\documentclass[twocolumn,aps,prl,showpacs]{revtex4}
\usepackage[dvips]{graphicx}
\begin{document}
\newcommand{\bea}{\begin{equation}}
\newcommand{\ber}{\begin{eqnarray}}
\newcommand{\eea}{\end{equation}}
\newcommand{\eer}{\end{eqnarray}}
\newcommand{\ct}{\cite}
\newcommand{\bi}{\bibitem}

\title{Dynamics of pulsed flow in an elastic tube}
\author{${}^{1}$Amit K. Chattopadhyay and ${}^{2}$Shreekantha Sil}
\address{{${}^{1}$Max Planck Institute for the Physics of Complex Systems,
N{\"o}thnitzer Stra{\ss}e 38, 01187 Dresden, Germany \\
${}^{2}$Department of Physics, Indian Institute of Science,
Bangalore 560012, India \footnote{Present address: Department of Physics, Viswa
Bharati University,
Santiniketan, West Bengal, India}}
\date{\today}
%
}
\begin{abstract}
Internal haemorrhage, often leading to cardio-vascular arrest 
happens to be one of the prime
sources of high fatality rates in mammals. We propose a simplistic
model of fluid flow to specify the location of the haemorrhagic
spots, which, if located accurately, could be operated upon leading to
a possible cure. The model we employ for the purpose is inspired by fluid
mechanics and consists of a viscous fluid,
pumped by a periodic force and flowing through an elastic tube. The 
analogy is with that of blood, pumped from the heart and flowing through an artery or vein. Our results,
aided by graphical illustrations, match reasonably well with experimental
observations. 
\end{abstract}
\maketitle

\section{Introduction}

The dynamics of fluid flow through elastic tubes occur in many natural
processes, like the organic respiration process, fluid flow in spongy bodies
and flow of blood in cardiac systems. Studies of these phenomena have covered a 
substantial portion of the literature of biological systems [1-15]. 
A typical example is the flow of blood through arteries and veins which are not rigid tubes but elastic tubes. 
\par
These real life systems have been the objects of investigation for quite a
while now and in many of these processes, nonlinear instabilities have been
observed to occupy positions of paramount importance \cite{alstrom,gustaffson},
 often something which can be
linked with the likes of Rayleigh or the "pearling" instability seen in tubular
lipid membranes \cite{barziv}. Presently, the major emphasis on cardiac related
studies seem to be on the electrophysiology of cardiac tissues during 
ischemic jolts \cite{fenton99} and attempts on linking different aperiodic 
cardiac processes to studies in chaos and turbulence 
\cite{sitabhra,narayanan,linhugh,stanley}. 
Other studies have focussed on the time series analysis of the
vibrating heart \cite{wessel,liebovitch,rappel,amaral98} with efforts on 
explaining the pulsating dynamics of cardiac muscles in the more physically 
quantifiable language of chaos and associated instabilities. 
Works in a somewhat different line have shown that applying fundamental
principles of nonlinear physics, it is possible to forecast beforehand the 
cardiac disarray arising due to rapid heart rates (ventricular tachycardia (VT))
\cite{wessel} or that due to aperiodic beating of the heart (ventricular
fibrillation (VF)) \cite{wessel,liebovitch}. Of the wide range of possibilities for their origin, researches have been focussed on the transition from
tachycardia to fibrillation, generating spiral waves 
\cite{fenton99,sitabhra,hastings}
to the anisotropic nature of ventricles \cite{fenton98} leading to faster 
electrical propagation parallel to the long axis of muscle fibers. The 
different methods of electrical defibrillation which have been presently
proposed \cite{sitabhra,panfilov,rappel} bear testimony to the intense efforts
on arriving at medical solutions from theoretical studies of bio-systems. In
this paper, we attempt to have an understanding of a complex
biological process, that of internal haemorrhage in an artery or vein,
through comparisons with an analogical mechanical model. 
It is to be noted that our approach is much more generalized in nature than 
that due to Sinha, {\it et al} \cite{sitabhra} or by Panfilov, {\it et al}
\cite{panfilov}, in the sense that we not only consider the case of 
ventricular fibrillation as a cause for aperiodicity in blood flow but we 
include all possible external disturbances (like a sudden thud on any portion
of the body) that might lead to internal haemorrhage. As such, our approach 
includes the dynamics of waves propagating around a steady obstacle 
\cite{xie,rappel} within the heart (clotted blood, in our case) as well as 
effectual VF. However, it needs to be mentioned that since all our
conclusions would be drawn via a mapping to a much simpler mechanical
system, we have no claims to exactness in terms of biological details.
But as we would indicate in the end, our qualitative
understanding can possibly be generalized to enable further
physiological details, through a simple extension of our model.  
In the following analysis, the main focus of our
study would be to determine the point of internal haemorrhage in a
single channel flow, using fundamental principles of fluid mechanics.
\par
This necessitates a clearer picture of the real life system in operation.
Experiments performed long ago by Brecher, {\it et al} \cite{brecher}, on the 
steady flux of blood through the superior vena cava of a dog with respect to 
the pressure difference in the jugular vein, although initially shows a linear 
behavior, the flux somewhat paradoxically attains a maximum value and no longer
increases. 
It is known for a long time that the flow of blood is pulsatile as a
consequence of the beating of the heart. The beating heart produces a pressure
pulse that travels through the blood, and this pressure wave is the pulse felt
in arteries. However there is a marked difference between the acoustic waves
generated in the heart and those which are felt in arteries. For the former,
the waves are a consequence of the compressibility of the blood and of the
living tissues surrounding it; while for the latter, the wave owes its
existence to the elasticity of the arteries and the coupling of the vibrations
of the artery walls to the blood flow. In 1961, Bergel \cite{bergel} studied the
velocity of propagation of the pulse wave through the thoracic aorta. In his 
derivation of the velocity of blood through arteries, he considered blood as
an inviscid and incompressible fluid. Inspite of this non-realistic assumption
of an inviscid fluid, which technically means that the fluid is supposed to be
in the high Reynolds' number zone, he obtained a wonderful match with
the experimental value of the velocity obtained by McDonald \cite{mcdonald2}. 
However the calculation predicted that the pulse travels undeformed during
propagation, and this is in sharp contrast to experimental observations. Real
life data show that the pulse wave changes its shape slowly as it travels
through the major arteries. This immediately suggests that to model a real
life system, it is extremely important to take into account the viscous nature of the blood. In this work,
we study the deformation of a pressure wave (the heart pulse) when it propagates through a
viscous fluid (the blood) flowing through an elastic tube. Our analysis reveals a wonderful
set of results inherent to the viscous nature of blood and which match with the
experimental observations. 

\section{The Theoretical Model}

The starting point of our analysis is the well-known Navier-Stokes' equation
in fluid mechanics \cite{landau} and the equation of continuity, for the 
case of a viscous, compressible flow
through an elastic tube. We consider an elastic tube filled with a fluid at
rest and surrounded by a fluid. The radius of the tube will be determined by the
transmural pressure difference between the interior and exterior pressures, as
well as the tension in the wall of the tube. Under normal circumstances, the
thickness of a blood vessel wall is small compared to the resting radius of
the blood vessel. Consequently, to a good approximation, we can treat the wall
as a thin membrane. Since the pulsating waveform causes the cross-sectional
area to depend on the spatial location, let A(x) be the cross-sectional area of the tube at a distance x, the pressure at that point, at time t be p(x,t) and the density is taken to be $\rho$. If u(x,t) be the velocity parallel to the tube axis, the continuity equation gives, 

\bea
\frac{\partial}{\partial t}(\rho A) + \frac{\partial}{\partial x}(\rho Au) = 0
\eea
 
\noindent
The modified Navier-Stokes' equation in the absence of a perpendicular component of the velocity reduces to, 

\ber
A[\frac{\partial}{\partial t}(\rho u) &+& u \frac{\partial}{\partial x}(\rho u)]
= -\frac{\partial}{\partial x} [(p-p_0)A] \nonumber \\
&+& \nu A \rho \frac{1}{r}
\frac{\partial}{\partial r} [r \frac{\partial u}{\partial r}] + \nu A \rho
\frac{{\partial}^2 u}{\partial {x}^2} 
\eer

\noindent
where r gives the radius of the tube at time t and $ \nu $ is the Newtonian
(and not the kinematic) viscosity of the fluid.
\par
Assuming linearized pressure dependence of density, {\it i. e.} $ p-p_0 =
{c_0}^2 (\rho-{\rho}_0) $ ($ c_0 $ being the mean velocity of blood) and
considering the deformable material of the walls of the blood vessels obey
Hooke's law, the pressure difference across the walls of a thin membrane
is given by \cite{rubinow2}, 
 
\bea
p(x) - p_0 = \frac{Eh}{r_0} (1-\frac{r_0}{r})
\eea

{\noindent}
where E is the Young's modulus of the walls of the artery (or vein), $ r_0 $ is the mean radius and h is the thickness of the artery (or vein) wall. Now we address the important question, that of the relative
importance of the nonlinear terms with respect to the linear ones in the above 
constitutive equations. A simple look at the magnitude of the Reynold's number 
(obtained from putting in experimental data \cite{bergel}) tells us that it 
has a value less than 60 \cite{ref1} and naturally we expect the dynamics to 
lie within the inertial zone. Physically speaking, incorporation of nonlinear 
terms would mean considering the effects of microscopic fluctuations occurring 
at the boundaries of the artery (or vein) walls due to the periodic input 
pulse. But the walls of the artery (or vein) being highly elastic (that is 
quite 
susceptible to stress), the average width of these fluctuations would be much 
smaller in magnitude than the mean radius $ r_0 $. Thus we can safely drop all 
nonlinearities henceforth. As an additional comment, it might be suggestive 
that this dropping of higher-ordered terms is only to aid an exact analytical
solution, although it is indeed a trivial exercise to exactly solve the two 
coupled equations numerically. However, as has been already indicated,  
this would not add to the {\em physics of mean-flow} in any way. 

Linearizing equations
(1) and (2) and neglecting all second-ordered terms (alternative linearization 
schemes in the contexts of reaction-diffusion systems, which can also be 
mapped to fluid flows, are to be found in \cite{hakim}), these respectively 
reduce to

\bea
{\rho}_0 \frac{\partial u}{\partial x} =  -\beta \frac{\partial p}{\partial t}
\eea

\noindent
where we use the fact that A=$ \pi r^2 $, 
 and $ \beta $ has the value $ \beta =
 \frac{1}{A_0}(\frac{{\rho}_0}{\alpha}+\frac{A_0}{{c_0}^2}) $, $ \alpha $ being equal to $ \frac{Eh}{2r_0 A_0} $. Eqn.(2) reduces to

\ber
A_0{\rho}_0 \frac{\partial u}{\partial t} &=& -A_0\frac{\partial p}{\partial x}
+ \nu A_0 {\rho}_0 \frac{{\partial}^2 u}{\partial r^2} + \frac{\nu {\rho}_0
A_0}{r} \frac{\partial u}{\partial r} \nonumber \\
& + & \nu {\rho}_0 A_0 \frac{{\partial}^2
u}{\partial x^2}
\eer

{\noindent}
It is to be noted that $ \beta $ is basically independent of $ c_0 $ once the 
latter becomes large (which it indeed is), relative to the other term. 

Now, considering the variation of the last equation with respect to
$x$ and utilising equation (4), we get

\begin{eqnarray}
\frac{\partial}{\partial t} (-\beta \frac{\partial p}{\partial t}) &=& 
-\frac{{\partial}^2 p}{\partial x^2} + \nu \frac{{\partial}^2}{\partial r^2}
(-\beta \frac{\partial p}{\partial t}) + \frac{\nu}{r} \frac{\partial}
{\partial r} (-\beta \frac{\partial p}{\partial t}) \nonumber \\
&-& \frac{\nu \rho_0}{r^2} \frac{\partial r}{\partial x} \frac{\partial u}
{\partial r} + \nu \frac{{\partial}^2}{\partial x^2} 
(-\beta \frac{\partial p}{\partial t})
\end{eqnarray}

\noindent
Rearranging and once again plugging in results from
equation (4), we finally obtain the complex wave equation as

\bea
\beta \frac{{\partial}^2 p}{\partial {t}^2} = \frac{{\partial}^2 p}{\partial {x}^2} + \gamma \frac{\partial p}{\partial t} + \lambda \frac{{\partial}^2}{\partial {x}^2} (\frac{\partial p}{\partial t})
\eea

\noindent
with $ \lambda = \nu \beta $ and $ \gamma = \frac{3\pi \lambda}{A_0}
$. All quantities with
suffix '0' denote the corresponding parameters for the unstretched elastic
tube. In the above derivation, spatial variations in density were ignored (not
exactly a {\it Boussinesq} approximation) to avoid
nonlinearities. More specifically, in arriving at equation (7) from
(6), we have used the relations $ \frac{\partial^2}{\partial r^2}(\frac{\partial
  p}{\partial t}) = \frac{6 \pi}{A} \frac{\partial p}{\partial t}
\approx \frac{6 \pi}{A_0} \frac{\partial p}{\partial t} $, $
\frac{1}{r} \frac{\partial}{\partial r} (\beta \frac{\partial
  p}{\partial t}) = -\frac{2}{r^2} \frac{\partial p}{\partial t}
\approx -\frac{2 \pi}{A_0} \frac{\partial p}{\partial t} $ and 
$ \frac{\rho_0}{r^2} \frac{\partial r}{\partial x} \frac{\partial
  u}{\partial r} = -\frac{\beta}{r^2} \frac{\partial p}{\partial t} 
\approx -\frac{\pi \beta}{A_0} \frac{\partial p}{\partial t} $. 
\par
The second and third terms appearing on the right hand side of the above
equation represent the damping rate due to viscosity. In fact, this dissipation is
responsible for the deformation of the wave pulse and is of paramount
importance in our theory. 
Now following usual methods, the modulated pressure wave is represented as

\bea
p_k = p_0\:e^{-\frac{(\lambda k^2-\gamma)}{2\beta}t}\:\times\:e^{
i(\frac{1}{2\beta} \sqrt{-(\gamma-\lambda k^2)^2 + 4\beta k^2}\:\:t + k x)}
\eea

\noindent
Damping of the wave occurs for those Fourier components for which $ k^2 \leq
\frac{\gamma}{\lambda} $. This implies that the Fourier modes with $ k_0 =
\sqrt{\frac{\gamma}{\lambda}} = \frac{\sqrt 3}{r_0} $ will propagate through
the system without any distortion with velocity $
u_{k_0}=\frac{1}{\sqrt{\beta}}$. Interestingly the velocity of this particular
mode matches with the results obtained by Bergel \cite{bergel}. The general 
dependence of velocity on k is given by

\bea
u_k = \frac{1}{\sqrt{\beta}}\:\sqrt{1-\frac{(k^2 \lambda-\gamma)^2 \beta}{4 k^2}}
\eea

\noindent
and this implies that the propagation mode exists only within the selected range

\bea
\sqrt{\frac{\gamma}{\lambda}} < |\vec k| < \frac{\sqrt{\beta}}{\lambda}
+ \frac{\sqrt{\beta + \gamma \lambda}}{\lambda}
\eea

\noindent
The key point in the above velocity expression is that it is, in
general, highly dispersive. However, the fact that we have arrived at
equation (8) utilising equations (4) and (7) precludes the possibility
that in the process of Fourier transforming the measured velocity to the wave
vector space, localized singularities would be lost and this means
that our analysis is practically valid only for highly dispersive
haemorrhages. This, though, is not a major deterrent, since almost all
real haemorrhagic conditions conform to such a regime. 

\section{Comparison with experimental data and results}

In the following we give a model estimation of the velocity of blood as calculated from
our theory utilizing parameters for blood flowing through the thoracic aorta
of dogs \cite{bergel}. Here $ E=4.3 \times {10}^6 $ dynes/$ {\mathrm{cm}}^2 $,
 $ h/r_0=0.105 $, $ {\rho}_0=1.06 $ gm/c.c., $ p_0=100 $ mm of Hg pressure, $
\nu=0.035 $ dyne-secs/$ {\mathrm{cm}}^2 $ , $ r_0=0.216 $ cm, $
c_0=1571 $ m/s. With these values, the undamped wave velocity ($ u_{k_0} =
\sqrt{\frac{\gamma}{\lambda}} $),
calculated from our theory (4.61 m/s), comes out very close to the 
mean velocity of blood flowing through the thoracic aorta of dogs ($
\approx $ 4.6 m/s) \cite{rubinow1,mcdonald2}. Although this is a
surprise agreement, considering the fact that our theory is valid only
for a single-channelled flow, while the experimental \cite{mcdonald2}
measurement is for the total multi-channelled network, we feel that
this happens because of the rather {\em insulated} nature of the
thoracic aorta compared to its neighborhoods, which in effect, 
probably renders it as some approximation of a single-channelled
tube. This seems an interesting comparison of our theory with
experiments, although, at this stage, we would not
dare any further conclusions based only on this information.
\par

\begin{figure}
\includegraphics[height=8.0cm,width=8.0cm,angle=-90]{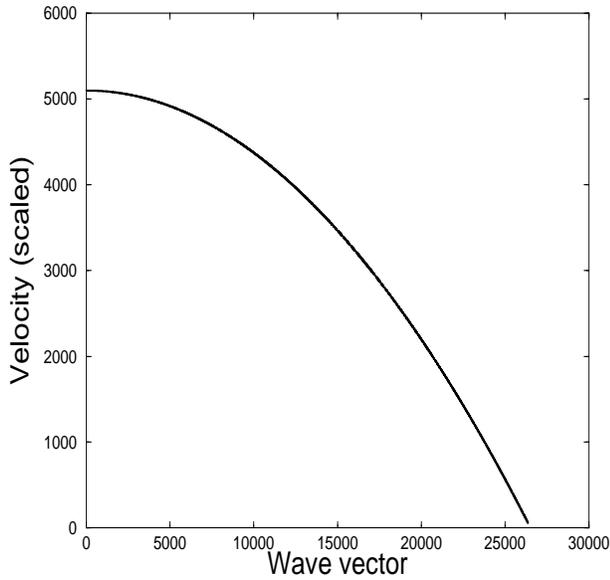}
\caption{Velocity vs wave vector plot, suitably scaled.}
\label{Fig:1}
\end{figure}

In Fig.I we show the variation of velocities (actually the scaled velocity
difference, to facilitate plotting) of different modes with respect
to the wave vector, where the corresponding parameters (mentioned above) are
used.

From this figure, it appears that
the velocity decreases with the increase of wave number, which is an obvious
signature of the presence of viscous damping in the system. This information,
although apparently a bit more on the mathematical side, in fact,
can be of crucial importance to the medical person in analysing the
exact location of the haemorrhagic spot. For this all that the doctor needs to 
know are the blood velocities at a few arbitrary spatial localions, in any 
major aorta or vein (say, pulmonary). The remaining exercise would be a 
technical triviality where a graph is to be drawn plotting blood velocities 
against the corresponding spatial locations and these data would then be 
translated into the wave-vector space through a computer calibration for exact 
comparison with our theoretical results. Now
whenever there is any aperiodicity in the blood flow, due to internal 
haemorrhagic clotting, the Fig.I. would show discontinuities in the form of 
spikes in the velocity spectrum, at the sites where
haemorrhage has occurred. In the language of physics, these are nothing but 
effects of varying initial conditions originating from the affected sites. The
fact that observation of the flow in any artery or vein is sufficient is
directly related to the fact that the arterial or venal system is a 
multi-connected network, which means that the discontinuity arising in one 
branch will carry through to all subsequent branches. A following inspection 
of the graph will immediately give an idea on where this 
spot is to be expected, or at least a rough idea of the zone to
operate on. Having said that, we should still be conscious of the fact
that our theory presented here is basically for a pulsed flow in a
{\em single} elastic tube. To have an exact comparison with the real
biological system, we need to consider the multi-channelled branching
of arteries (or veins), which certainly is a more complex network. The
importance of our theory lies in the fact that to have an
understanding of this multi-channelled network, we need to know the
dynamics of each individual channel, after which an analysis of the apparently
complicated network boils down to a simple enough initial-value problem at
each individual node, something which is easily solvable numerically. 

\par
In the following graphical illustrations, we consider two different
initial conditions and follow their evolution with time as defined in eqn.(7), 
in a co-moving frame of reference.
\begin{figure}
\includegraphics[height=8.0cm,width=8.0cm,angle=-90]{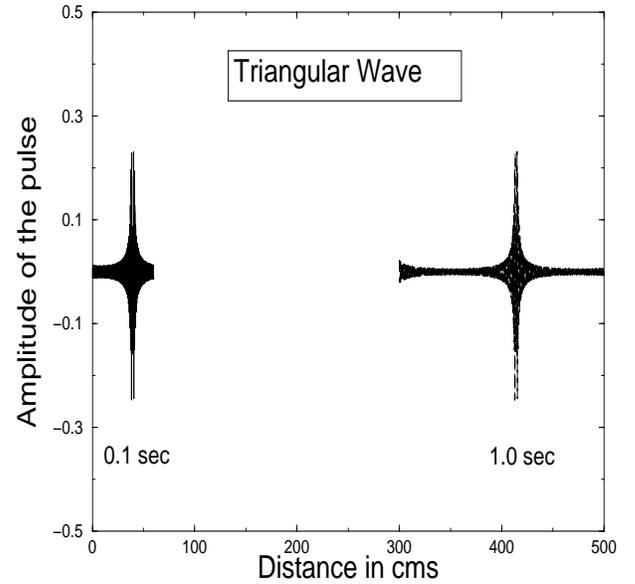}
\caption{Comparison between triangular wave pattern vs distance graphs after 
0.1 sec and 1.0 sec.}
\label{Fig:2}
\end{figure}

Fig.II gives the
comparison between the spatial variation of an input hypothetical triangular
wave at times 0.1 sec and 1 sec. These while obeying the same dynamics as wave
pulses through an elastic tube, clearly show damping along the line of
propagation. 

\begin{figure}
\includegraphics[height=8.0cm,width=8.0cm,angle=-90]{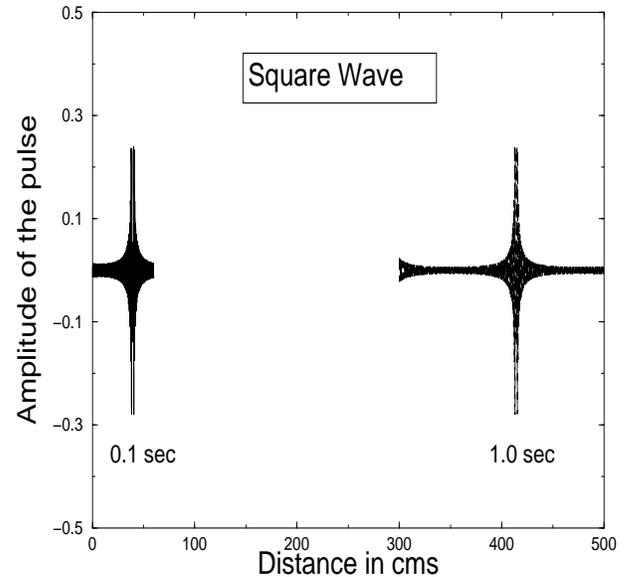}
\caption{Comparison between square wave pattern vs distance graphs after 
0.1 sec and 1 sec.}
\label{Fig:3}
\end{figure}

Fig.III shows identical variations for an input square
wave-train (hypothetical) at the same times previously mentioned. 

Finally, a comparison between two completely different hypothetical input 
pulses has been given in Fig.IV at 0.1 sec. which shows
that although the pulses might start with widely varying initial conditions,
the damping mechanism operating with non-zero viscosity leaves very little
difference between their output waveforms after times of the order of 0.1 sec 
(and larger times).

\begin{figure}
\includegraphics[height=8.0cm,width=8.0cm,angle=-90]{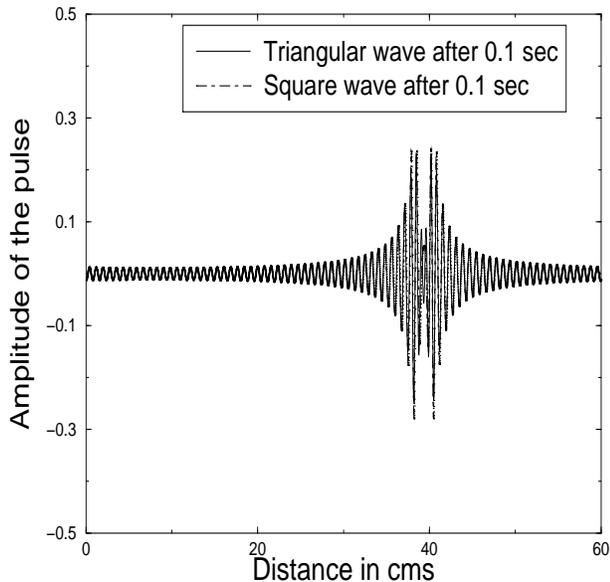}
\caption{Comparison between triangular and square wave pattern vs distance
graphs after 0.1 sec. The two graphs are almost indistinguishable.}
\label{Fig:4}
\end{figure}

This study leads us to the conclusion that inspite of starting with widely
varying initial conditions, the pressure pulses lose their memory very quickly
and develop exactly identical forms. This happens because except a
particular mode, the critical mode $ k_0 =
\sqrt{\frac{\gamma}{\lambda}} $, in the Fourier spectrum, all other
modes have non-vanishing attenuation. However, for small time
evolution, in the scales of measurements ($\sim$ millisecs), the
patterns are perfectly distinguishable and may be utilized in getting
an idea of the variation of the mean viscosity of the blood to the
local viscosity corresponding to a particular wave vector mode, from a 
knowledge of the critical mode.

\section{Conclusions and Discussions}

Our objective was to devise a theory which is capable, more or 
less exactly, to pinpoint the location of an internal haemorrhagic
spot, in a single arterial or venal channel, 
due to clotting of blood. A knowledge of this 
can easily be obtained from Fig.I by comparing the variations of the velocity
of the flow against the spatial locations at different points of any artery or 
vein, more technically speaking, by looking at the eventual discontinuities in 
the spectrum. 
The fact, that the attenuated waveforms become independent of the
nature of the input waveforms, after times of the order of 0.1 sec, is
proved by Figs.II-IV. This fact is of immense importance to our description,
since, this in a way points to sort of an {\em universality} in the nature of 
our analysis. To be more specific, our analysis shows that we do not even need 
an exact description of the nature of the arterial or venal pulses in 
individuals to apply our theory. Rather the nature of the arterial (or venal) 
dynamics adjusts itself in such a self-consistent way that all specific input 
details are washed out and only the system elastic properties become the sole 
deciding parameter of the position of eventual blood clotting. The
propagating pressure waves are defined only within a wave-vector window
(eqn.(9)) and avoid possibilities of unwanted instabilities in
eqn.(7). Incidentally, it might be noted that our theory is exact for blood
plasma, which is a very close approximation to a Newtonian fluid, although,
blood, in general has minor non-Newtonian characteristics. 
As results obtained from the model
calculations show that our observation tallies extremely well with the 
experimental value of the velocity of blood flowing through the thoracic 
aorta of dogs, we have justified hopes that compared with real life 
situations, the location of the haemorrhagic spot, calculated from an
extended version of our theory and including the multi-channelled
branching of the arterial (or venal) network, would lie within
acceptable error limits. However, as is the wont of any basic theory, ours 
too is wrought with a few characteristic assumptions as we have indicated 
earlier too. The theory presented here is certainly not complete enough for
comparison with a real life experiment done on a multi-connected
arterial (or venal) network, not at least at this stage. But our objective was
basically to describe the dynamics of flow in each individual
channel of the multi-connected network, which, with well defined
initial conditions, could simulate the entire description of a real
life flow. We look forward to experimental verifications of our theory, with
experiments done on a single tube and designed essentially in the way we have
described earlier in the text. For more realistic comparisons, we are working
on a numerical model with all the multi-channelled subtleties involved
\cite{amit}, an essential coupling of single-channeled information analysed here
to the generation of a multi-channelled network.
\par
AKC acknowledges discussions with J. K. Bhattacharjee and C. Dasgupta 
during this work.

\begin{references}
\bibitem{ref1} Reynold's number, $ Re = \frac{v L}{\nu/\rho} $, where $ v $ is
the velocity of the flow and $ L $ is the length scale of the geometry of the 
flow, $ \nu/\rho $ being the kinematic viscosity. For blood flowing through 
the aorta of a human body, $ v \sim $ 30 cm/secs. \cite{folkow} and from Fig.IV
we see that even at the order of 0.1 sec, the solutions remain initial 
condition independent. So we chose t=0.001 sec and putting in the values of 
the parameters, it can be easily shown that $ Re < 60 $.
\end{references}
\end{document}